\documentstyle[prl,multicol,epsf,aps]{revtex}
\begin{document}
\epsfverbosetrue
\title{Defect modes in one-dimensional photonic lattices}
\author{Francesco Fedele, Jianke Yang \\
Department of Mathematics and Statistics, University of Vermont,
Burlington, VT 05401 \\
Zhigang Chen \\ 
Department of Physics and Astronomy, San Francisco State University, 
San Francisco, CA 94132 }
\maketitle

\begin{abstract}
Linear defect modes in one-dimensional photonic lattices are studied theoretically. 
For negative (repulsive) defects, various localized defect modes are found. 
The strongest confinement of the defect modes appear when the
lattice intensity at the defect site is {\em non-zero} rather than zero. 
When launched at small angles into such a defect site of the lattice, a Gaussian beam can be trapped and undergo snake oscillations under appropriate conditions. 
{\it OCIS numbers: 190.5530, 230.3990.}
\end{abstract}

\begin{multicols}{2}
\narrowtext
\vspace{0.5cm}

Light propagation in periodic photonic lattices is under intensive study these days due to their novel physics and light-routing applications \cite{Dmitrie_Nature,Kivshar_Review}. 
Most of these studies focused on nonlinear light behaviors in uniformly periodic 
lattices \cite{Aceves,Eisenberg,Fleischer03,Kivshar,Chen,Yang,Stegeman,Torner}. A natural question arises: how does light propagate if the photonic lattice has 
a local defect? In photonic crystals, this question has been much analyzed
\cite{Joannopoulos}.  
While nonuniform arrays of "fabricated" waveguides with structured defects
were used in previous studies \cite{defect1,defect_kivshar,defect2}, the issue of defect modes in optically-induced photonic lattices has not yet received much attention. Since photonic lattices
differ from photonic crystals on many aspects (for instance, the refractive-index variation in an induced photonic lattice is typically several orders of magnitude smaller than that in a photonic crystal), one wonders if photonic lattices with a local defect can also support defect modes. 

In this Letter, we theoretically
analyze linear defect modes in one-dimensional photonic lattices
with a local negative defect as induced in a biased photorefractive crystal. In such a defect, the lattice intensity
is lower than that at nearby sites (akin to an "air defect" in photonic crystals \cite{Joannopoulos}), thus light has a tendency to escape from
the defect to nearby sites. 
However, we found that localized defect modes do exist due to repeated Bragg reflections. 
More interestingly, strongly confined defect modes appear when the 
lattice intensity at the defect site is {\em non-zero} rather than zero.
As the lattice potential increases (by raising the bias field), 
defect modes move from lower bandgaps to higher ones. 
If a Gaussian beam is launched at small angles into the defect, it can be trapped and undergo robust snake oscillations inside the defect site without much radiation. 

The physical situation we consider here is that an ordinarily polarized lattice beam with a single-site negative defect is launched into a photorefractive crystal. This
defected lattice beam is assumed to be uniform along the direction of propagation. Meanwhile, an extra-ordinarily polarized probe beam with a very low intensity is launched into the defect site, propagating collinearly with the lattice beam. The non-dimensionalized model equation for the probe beam is 
\cite{Fleischer03}
\begin{equation} \label{linear_model}
iU_z+U_{xx}-\frac{E_0}{1+I_{L}(x)}U=0. 
\end{equation}
Here $U$ is the slowly-varying amplitude of the probe beam, 
$z$ is the propagation distance (in units of
$2k_1D^2/\pi^2$), $x$ is the transverse distance (in units of $D/\pi$), 
$E_0$ is the applied DC field [in units of $\pi^2/(k_0^2n_e^4D^2r_{33})$],  
$I_L=I_0\cos^2x\left\{1+\epsilon f_D(x)\right\}$ 
is the intensity function of the photorefractive lattice (normalized by
the dark irradiance of the crystal $I_d$), $I_0$ is the peak intensity of 
the otherwise uniform photonic lattice (i.e., far away from the defect site), 
$f_D(x)$ is a localized function describing the shape of the defect, 
$\epsilon$ controls the strength of the defect, 
$D$ is the lattice spacing, $k_0=2\pi/\lambda_0$ is 
the wavenumber ($\lambda_0$ is the wavelength), $k_1=k_0 n_e$,  $n_e$ 
is the unperturbed refractive index, and
$r_{33}$ is the electro-optic coefficient of the crystal.
In this letter, we assume that the defect is restricted to a single
lattice site at $x=0$. Thus, we take $f_D(x)=exp(-x^8/128)$. 
Other choices of defect functions $f_D$ give similar results. When 
$\epsilon<0$, the light intensity $I_L$ at the defect site is lower 
than that at the surrounding sites. This is called a negative (repulsive) defect
where light tends to escape to nearby lattice sites. For $\epsilon=-1$ and $-0.5$, 
the corresponding lattice intensity profiles are displayed in Figs. 1a and 3b respectively. In the former case, there is no light at the defect site, 
while in the latter case, there is still light at the defect site but with a halfway reduced intensity. These lattices with structured defects might be generated experimentally by optical induction. 
Consistent with our 
previous experiments \cite{Chen}, we choose parameters as follows: the lattice intensity $I_0=3I_d$, 
lattice spacing $D=20\mu$m, $\lambda_0=0.5\mu$m, $n_e=2.3$, and $r_{33}=280$pm/V.
Then one $x$ unit corresponds to $6.4 \mu$m, 
one $z$ unit corresponds to 2.3 mm, and one $E_0$ unit corresponds
to 20 V/mm in physical units. 

For a negative defect, a surprising feature is the possible existence of 
localized defect modes due to repeated Bragg reflections. The existence
of such modes will affect light propagation in a profound way. 
We seek such modes in the form 
$U(x, z)=e^{-i\mu z} u(x)$, 
where function $u(x)$ is localized in $x$, 
and $\mu$ is a propagation constant. 
Our numerical method is to expand the solution $u(x)$ into discrete Fourier series, 
then converting the linear $u(x)$ equation into an eigenvalue problem with $\mu$ as
the eigenvalue. 
First, we consider the defect with $\epsilon=-1$, 
where the lattice intensity at the defect is zero (see Fig. 1a). 
For this defect, we have found defect modes at various values of $E_0$. 
The results are shown in Fig. 1b. It is seen that at low values of $E_0$
(low potential), two defect modes appear in the first and second bandgaps. 
The one in the first bandgap is symmetric in $x$, while the one
in the second bandgap is anti-symmetric in $x$. Both types of defect modes are moderately confined. Examples of such modes at $E_0=1.2$ and 3.5 are displayed
in Fig. 1(c, d) respectively. 
However, these defect modes disappear when $E_0$ increases above certain
threshold values. In particular, the symmetric branch in the first bandgap
disappears when $E_0>2.8$, while the anti-symmetric branch in the second
bandgap disappears when $E_0>7.5$. 
On the other hand, before the antisymmetric branch disappears, 
another symmetric branch of defect modes appears inside the same (second) bandgap. This 
new branch exists when $5.3<E_0<10.3$, and it is generally 
more localized than the previous two branches. 
This can be seen in Fig. 1e where this symmetric defect mode
at $E_0=7.5$ is illustrated. Compared to Fig. 1(c, d), this new mode is much more confined. 

Three general features in Fig. 1(b) should be noted. First, 
for any positive $E_0$ value, at least one defect mode can be found. 
Second, each branch of defect modes disappears as $E_0$ increases to above
a certain threshold. 
Thirdly, as $E_0$ increases, defect modes disappear from lower bandgaps
and appear in higher bandgaps. In other words, defect modes move from lower bandgaps to 
higher ones as $E_0$ increases. 

The existence of these defect modes as well as their profile and
symmetry properties have a profound effect on linear light propagation
in the underlying defected photonic lattices. If the input probe beam
takes the profile of a defect mode, then it will propagate stationarily
and not diffract at all. This is seen in Fig. 2(b), where the numerical 
evolution of an initial defect mode (with $\epsilon=-1$ and $E_0=7.5$)
is displayed (the corresponding lattice field is shown in Fig. 2a). 
For a Gaussian input beam (as is customary in experimental conditions), 
the evolution will critically depend on whether a defect mode resembling
the input Gaussian beam exists under the same physical conditions. 
To demonstrate, we take an initial Gaussian beam as 
$U(x, 0)=e^{-\frac{1}{3}x^2}$
which resembles the central hump of the defect mode in Fig. 1e, 
and simulate its evolution under various $E_0$ values by pseudo-spectral methods. 
The lattice intensity field is the same as that in Fig. 2a (where $\epsilon=-1$). 
We found that at small values of $E_0$, the Gaussian beam strongly 
diffracts and quickly becomes invisible. Similar behavior persists
as $E_0$ increases (see Fig. 2c) until it reaches a value about 7.5, 
when a large portion of the initial beam's energy is trapped inside the 
defect site and propagates stationarily (see Fig. 2d). 
As $E_0$ increases beyond 7.5, however, strong diffraction of the probe is seen again
(see Fig. 2e). 
These results indicate that the light trapping in Fig. 2d 
could not be attributed to either the simple guidance due to increased lattice potential or the nonlinear self-action of the probe beam itself. 
Rather it must be attributed to the repeated Bragg reflections inside the 
photonic lattice under certain phase-matching conditions, as the Gaussion beam matches the localized mode of the defect. This bears strong resemblance to localized modes in photonic crystal fibers.

For various applications, it is often desirable to keep the defect modes as locally confined 
as possible. The defect considered above with $\epsilon=-1$ 
(see Figs. 1a and 2a) is certainly simple and
intuitive, but does it give the most strongly confined defect modes? To answer this question, 
we fix the value of $E_0$  and allow the defect parameter $\epsilon$ to vary from $-1$
to $0$, then determine at what $\epsilon$ values the most localized defect modes arise. 
With fixed $E_0=6$, we have obtained the defect modes versus $\epsilon$ and
plotted the results in Fig. 3. 
Fig. 3(a) reveals that at small negative values of $\epsilon$, a single
defect mode bifurcates from an edge of a Bloch band inside each bandgap. 
As $\epsilon$ decreases, 
the defect mode in the first bandgap disappears (at $\epsilon=-0.81$), 
while the one in the second bandgap persists. 
The defect-mode branch in the first bandgap is more localized than the one in
the second bandgap in general. Thus we focus on this branch in the first bandgap below. 
When $|\epsilon|$ is small, the defect eigenvalue is rather close to the
left Bloch band, thus the defect mode is rather weakly confined (see Fig. 3c). 
As $|\epsilon|$ increases, the mode becomes more confined. 
As $\epsilon$ approaches $-0.81$, the defect eigenvalue approaches the 
right Bloch band, and the defect mode becomes less confined again (see Fig. 3e). 
Surprisingly, we found that the strongly confined defect mode occurs when $\epsilon \approx -0.5$. 
This defect mode and the corresponding lattice intensity field are shown in 
Fig. 3(d, b) respectively. These findings are rather interesting, as they
show that the most localized defect mode arises when the lattice intensity
at the defect site is {\em non-zero} rather than zero. Such results may have
important implications for applications of defect modes in
photonic lattices. 

We have further studied the evolution of a Gaussian input beam launched at small angles
into a photonic lattice with $E_0=6$ and $\epsilon=-0.5$. For this purpose, 
we take the initial condition as 
$U(x, 0)=e^{-\frac{1}{3}x^2+ikx}$, 
where this Gaussian intensity profile resembles the central hump in the defect mode
of Fig. 3d, and the phase gradient $k$ is proportional to the launch angle of the Gaussian 
beam. At zero launch angle ($k=0$), a vast majority of the input-beam's energy is trapped
inside the defect and propagates stationarily (see Fig. 4b). 
When compared to Fig. 2, we see that the confinement of the probe beam by the present defect (shown in Fig. 4a) is more efficient, mainly because
the defect mode admitted under these conditions is more localized
(see Fig. 3d). 
Next we take $k=1$, which corresponds to a launch angle of $0.58^\circ$ with physical parameters listed earlier. 
In this case, most of the light is still trapped inside the defect site. However, the trapped light undergoes robust snake-like oscillations
as it propagates through the defect (see Fig. 4c). The ability of a negative defect to trap oscillating light beams is a remarkable feature that merits further investigation. 

In summary, we have analyzed linear defect modes in one-dimensional
photonic lattices with negative local defects. 
These results are expected to pave the way for experimental
observations of such localized modes as well as for the study of nonlinear defect modes. 

This work was supported in part by AFOSR, NASA EPSCoR grants, and ARO. J. Yang's email address is jyang@math.uvm.edu. 

\begin{figure}
\begin{center}
\setlength{\epsfxsize}{8.5cm} \epsfbox{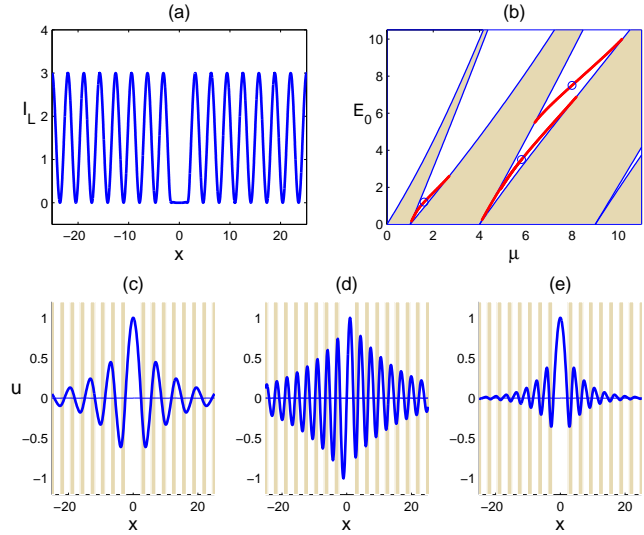}
\vspace{0.4cm}
\caption{(a) The lattice intensity profile with $I_0=3$ and $\epsilon=-1$; 
(b) the applied dc field parameter $E_0$ versus the defect eigenvalues $\mu$;  
the shaded regions are Bloch bands; 
(c, d, e) three defect modes at $(E_0, \mu)=(1.2, 1.604), (3.5, 5.812), (7.5, 7.997)$
which are marked by circles in (b) respectively. The shaded stripes indicate the
locations of the lattice's peak intensities. }
\end{center}
\end{figure}

\vspace{-0.4cm}
\begin{figure}
\begin{center}
\setlength{\epsfxsize}{8.5cm} \epsfbox{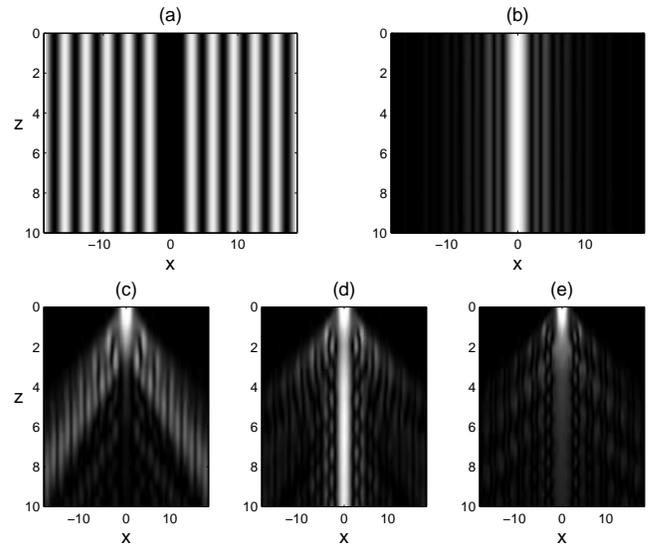}
\caption{(a) The lattice intensity field with $I_0=3$ and $\epsilon=-1$; 
(b) evolution of an exact defect mode (shown in Fig. 1e) at $E_0=7.5$; 
(c, d, e) evolutions of a Gaussian beam at three 
$E_0$ values $5, 7.5$ and 10 respectively.}
\end{center}
\end{figure}

\vspace{-0.7cm}
\begin{figure}
\begin{center}
\setlength{\epsfxsize}{8.5cm} \epsfbox{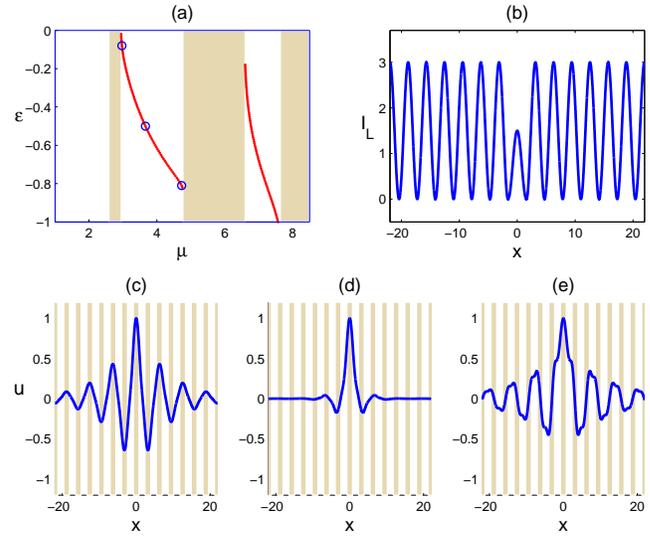}

\vspace{0.4cm}
\caption{(a) The defect strength $\epsilon$ versus the 
defect eigenvalues $\mu$;  
(b) intensity profile $I_L(x)$ of the photonic lattice with $\epsilon=-0.5$; 
(c, d, e) three defect modes of the first bandgap with 
$(\epsilon, \mu)$ as marked 
by circles in (a) respectively.  }
\end{center}
\end{figure}

\vspace{-0.4cm}
\begin{figure}
\begin{center}
\setlength{\epsfxsize}{8.5cm} \epsfbox{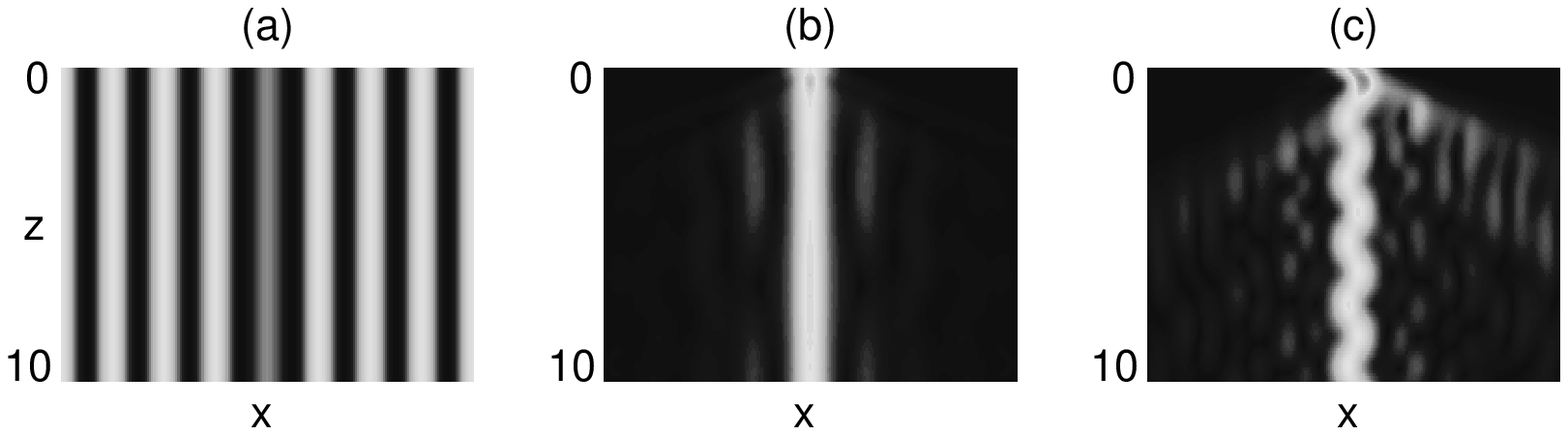}

\vspace{0.4cm}
\caption{Evolution of a Gaussian beam launched at zero (b) and non-zero (c) angles 
into the defect site of a photonic lattice (a). Intensity fields are shown. Here
$I_0=3, E_0=6$ and $\epsilon=-0.5$ in Eq. (\ref{linear_model}). The initial phase gradient
in (c) is $k=1$. }
\end{center}
\end{figure}

\end{multicols}
\end{document}